\documentclass{osa-article}

\journal{oe}

\begin{document}

\title{Loophole-free plug-and-play quantum key distribution}

\author{Ruoyang Qi,\authormark{1,2,7} Haoran Zhang,\authormark{1,7} Jiancun Gao,\authormark{1,3} Liuguo Yin,\authormark{3,4,5,6,8}and Gui-Lu Long\authormark{1,3,5,6,*}}

\address{\authormark{1}State Key Laboratory of Low-Dimensional Quantum Physics and Department of Physics, Tsinghua University, Beijing 100084, China\\
\authormark{2}Beijing Institute of Spacecraft System Engineering, Beijing 100094, China\\
\authormark{3}Frontier Science Center for Quantum Information, Beijing 100084, China\\
\authormark{4}School of Information and Technology, Tsinghua University, Beijing 100084, China\\
\authormark{5}Beijing National Research Center for Information Science and Technology, Beijing 100084, China\\
\authormark{6}Beijing Academy of Quantum Information Sciences, Beijing 100193, China\\
\authormark{7}These authors contributed equally to this work.\\
\authormark{8}yinlg@tsinghua.edu.cn

}

\email{\authormark{*} gllong@tsinghua.edu.cn} 



\begin{abstract}
Robust, simple, and flexible quantum key distribution (QKD) is vital for realising practical applications of this technique. 
Contrary to typical phase-coded QKD schemes, the plug-and-play QKD design requires only one arm-length-insensitive interferometer without active feedback, in which the noise is automatically compensated by the two-way structure. 
However, there are certain possible loopholes in the typical plug-and-play designs, which require consideration and strict monitoring. 
This study proposes a theoretically loophole-free plug-and-play QKD scheme with two-way protocol and presents an experimental demonstration of said scheme. 
The security is analysed under a collective attack scenario assisted by the decoy state method. 
The scheme was implemented in a 50.4 km commercial fibre without active feedback. 
The system showed highly robust performance with an ultra-low error rate and maintained an ultra-high visibility of $0.9947\pm 0.0002$ through significant environmental changes over 24 hours.
\end{abstract}

\section{Introduction}
The principles of quantum mechanics enable novel information security methods. Since the first quantum communication protocol was proposed in 1984 \cite{bennett2014quantum}, there have been rapid developments in both theory and experiments\cite{long2002theoretically,gisin2002quantum,bostrom2002deterministic,scarani2009security,
liao2017satellite,boaron2018secure}.
Quantum key distribution (QKD), a forerunner technique in the application of quantum information, can operate in optical fibre \cite{boaron2018secure} or in free space \cite{liao2017satellite}.
The signal can be encoded in various degrees of freedom, such as the polarisation and phase states of photons. Among them, phase and time-bin encoding \cite{brendel1999pulsed,boaron2018secure} are widely adopted in fibre systems due to their high transmission stability.

However, environmental noise presents a significant challenge for the practical application of QKD. The main factors are the mismatching and phase drift of the interferometers, and birefringence in the fibre. In particular, it is impossible to keep the relative length of the double Mach–Zehnder interferometer stable when the two legitimate users are separated \cite{muller1997plug}.
A widely adopted way to solve this problem is to use a pair of customised interferometers with a feedback system between the users, but it is not applicable in all situations. In addition to the complexity of implementing the system, rapid temperature changes and mechanical vibrations present challenges to the reliability and efficiency of the feedback system in practical applications, such as fibre hanging from a pole on a windy day, or other environmental conditions at the user’s locations. In addition, users need to share interferometers of the same model in advance, which limits the flexibility of the technique in some applications. Moreover, feedback between the users causes a much more real-time classical information interaction, which may lead to a greater risk of being hacked through side channels. Another method is to use the design of Martilelli \cite{martinelli1989universal}, in which the polarisation drift is automatically corrected through the use of a Faraday mirror with a two-way structure. Therefore, the plug-and-play scheme \cite{muller1997plug} can greatly reduce the noise without active feedback because the two pulses in the phase coding go through the same forward-backward path and compensate the polarisation fluctuation and phase drift of each other. However, there are various possible loopholes \cite{2006Phase,Gisin2006Trojan,2015Practical,Feihu2010Experimental,zhao2010security,2011Passive,xu2010passive}, such as phase-remapping attacks, Trojan horse attacks, and untrusted source attacks, which must be considered and strictly monitored. Typical two-way quantum communication schemes \cite{deng2004secure,deng2004bidirectional,lucamarini2005secure} can be proved to be unconditionally secure \cite{lu2011unconditional,beaudry2013security,henao2015practical,2015Finite,qi2019implementation}, but those photons chosen for monitoring go through only the forward channel, where the tailor-made interferometers and feedback system are still required.

Here, we propose a theoretically loophole-free plug-and-play scheme with a two-way QKD protocol based on non-orthogonal time-bin and phase states. When a pulse enters an asymmetric interferometer, it is split into two subpulses, where the one taking the short path is denoted as  $\left|0\right\rangle$, and that taking the long path is denoted as $\left|1\right\rangle$.
The time-bin states $\{\left|0\right\rangle,\left|1\right\rangle\}$ are the eigenstates of $\sigma_Z$.
The superposed phase states $\left|\pm\right\rangle=(\left|0\right\rangle\pm\left|1\right\rangle)/\sqrt{2}$ are the eigenstates of $\sigma_X$. 
The key principle is to encode bits of the key only in the states in the X-basis, and use the states in the Z-basis in the forward path to monitor the phase error of the states in the X-basis. Because the states in the Z-basis are highly stable against flip errors, neither an interferometer nor compensation of the phase or polarisation is required. To eavesdrop, Eve has to perform some operation over the forward channel \cite{lu2011unconditional}, causing errors in the Z-basis and eventually leading to their exposure. Because those bits carrying phase-encoded states pass through the same interferometer, both the forward and backward paths are automatically compensated and highly robust. In principle, the proposed design could find an entirely accurate measurement basis itself. In addition to the electronic control errors and constant background noise such as dark counts, the errors from mismatching, drift, and flip are almost non-existent. Even if the interferometer and the quantum channel are subject to significant environmental changes over an extended period of time, active feedback between the users is still unnecessary, which indicates the high robustness of this plug-and-play system.
\section{Results}
\subsection{The Protocol}
As is common in QKD, the public announcements by Alice and Bob are performed over an authenticated channel. For simplicity, we first assume that the source is an ideal single-photon source. The protocol contains the following four steps.
\begin{table}[htbp]
	\centering
	\caption{\bf Encoding of the states by Bob}
	\begin{tabular}{c c c c c}
		\hline
		\hline
		~ & ~~$\alpha_i$~~ & ~~$\beta_i$~~ & ~~$\xi_i$~~ & ~~State~~\\
		\hline
		{~Monitoring~} & 0 & 0 & 0 & $\left|0\right\rangle$\\
		\cline{2-5} ~ & 1 & 0 & $\pi /2$ & $\left|1\right\rangle$\\
		\hline
		{~Information carrier~} & 0 & 1 & $\pi /4$ & $(\left|0\right\rangle + \left|1\right\rangle)/\sqrt{2}$\\
		\cline{2-5} ~ & 1 & 1 & $-\pi /4$ & $(\left|0\right\rangle - \left|1\right\rangle)/\sqrt{2}$\\
		\hline
		\hline
	\end{tabular}
	\label{encoding}
\end{table}

\begin{enumerate}
	\item {\it State preparation:} Bob generates a sequence of single time-bin photon pulses for each time window $i$,
    \begin{eqnarray}
       \left| {{\psi _i}} \right\rangle  = \cos {\xi _i}\left| 0 \right\rangle  + \sin {\xi _i}\left| 1 \right\rangle,
    \end{eqnarray}
    where ${\xi _i} = \pi [1 - {\beta _i} - {( - 1)^{{\alpha _i} + {\beta _i}}}]/4$.
    As seen in Table \ref{encoding}, when $\beta_i = 0$, $\left| {{\psi _i}} \right\rangle$ is the eigenstates of $\sigma_Z$.
    When $\beta_i = 1$, $\left| {{\psi _i}} \right\rangle$ are the eigenstates of $\sigma_X$.
    Bob prepares the state in eigenstates of $\sigma_Z$ with probability $q$ and in eigenstates of $\sigma_X$ with probability $1-q$.
    Both the signal carrier states and the monitoring states are randomly in either $\alpha_i = 0$ or $\alpha_i = 1$.
    Then Bob sends the sequence of states to Alice, over the forward channel.
    The values of $ \alpha_i $ and $ \beta_i $ are never publicly announced.

\item {\it Error-check and encoding:}
	Alice measures the received states with probability $p$ in the Z-basis and sends the measured result and the position to Bob for estimating the bit error rate (BER) $e_m$, from those instances in which both Alice and Bob choose the Z-basis. 
	For the remaining states, Alice randomly encodes the bit value with two unitary operations $I = \left| 0 \right\rangle \left\langle 0 \right| + \left| 1 \right\rangle \left\langle 1 \right|$ and $\sigma_Z = \left| 0 \right\rangle \left\langle 0 \right| - \left| 1 \right\rangle \left\langle 1 \right|$, mapping to bit values 0 and 1, respectively.
	She sends the operated states back to Bob.

\item {\it Measurement and sifting:}
	Bob measures the states $\{\left|+\right\rangle=(\left|0\right\rangle+\left|1\right\rangle)/\sqrt{2},\left|-\right\rangle=(\left|0\right\rangle-\left|1\right\rangle)/\sqrt{2}\}$, in the X-basis.
	He only keeps the results for those he prepared in the X-basis, namely $\beta _i=1$.
	Bob discards the results if $\beta _i=0$.
\item {\it Post-processing:}
Bob announces the BER.
With the BER and these remain results, Alice and Bob can bound the information leakage and perform the post-processing such as error correction and privacy amplification\cite{fung2010practical}.
\end{enumerate}

\subsection{Security Analysis}
We restrict ourselves to the case of collective attack with an infinite scenario \cite{scarani2009security}.
According to Devetak-Winter’s theory \cite{csiszar1978broadcast,devetak2005distillation}, the maximum achievable secure key rate (SKR) is
\begin{eqnarray}
	R_s=I_{AB}-I_E,
\end{eqnarray}
where $I_{AB}$ represents the mutual information between Alice and Bob, and $I_E$ is the maximum amount of information an eavesdropper can obtain using the best possible strategy. 

The channel between Alice and Bob can be treated as a cascaded channel consisting of a binary symmetric channel and a binary erasure channel in series.
The supreme of the mutual information between Alice and Bob $I_{AB}$ is limited by the Shannon limit according to noisy-channel coding theorem,
\begin{eqnarray}
   I_{AB} = Q\cdot [1-f \cdot h(e)],
\end{eqnarray}
where $e$ is the BER between the legitimate users, $Q$ represents the probability of Alice obtaining a raw key bit and $f$ represents efficiency of error correction, and $h(x) =  - x{\log _2}x - (1 - x){\log _2}(1 - x)$  represents the binary Shannon entropy.
Some error correction codes such as low density parity check (LDPC) codes can approach very closely the Shannon limit.

Weak coherence pulses are often used as source instead of single photons, multi-photon pulses will enable Eve to obtain more information.
In such a case, the maximum amount of information Eve can acquire becomes \cite{gottesman2004security},
\begin{eqnarray}
   I_E = Q \cdot [\Delta_1 \cdot I_{E,1}+1-\Delta_1 -\Delta_0],
\end{eqnarray}
where $I_{E,1}$ represents the maximum amount of information of Eve can obtain from single photons, $\Delta_1$ and $\Delta_0$ are the fraction of detection events from single-photon and vacuum states, respectively, which can be accurately estimated by the decoy state method for the states in X-basis at Bob's end \cite{hwang2003quantum,wang2005beating,lo2005decoy}.

In the following, we restrict the upper bound of $I_{E,1}$.
The prepared quantum state is $\rho$ and we consider the case of collective attack, where the most general quantum operation Eve may perform in the forward channel consists of a joint operation on the qubit and some ancilla of Eve,
\begin{eqnarray}
{\rho ^E} = U( {\rho  \otimes \left| \varepsilon  \right\rangle \left\langle \varepsilon  \right|}){U^ + },
\end{eqnarray}
where $\left| \varepsilon  \right\rangle$ represents Eve's ancillary state and $U$ is a unitary operation acting on the joint space of the ancilla and the qubit. 
After the operation performed by Alice, the state becomes,
\begin{eqnarray}
{\rho ^{AE}} = \frac{{\rho _0^E + \rho _1^E}}{2},
\end{eqnarray}
where $\rho _0^E = I\rho^E {I^+}$ and $\rho _1^E = \sigma_Z\rho^E {\sigma_Z^+}$ are the encoded states of $0$ and $1$, respectively.
The information Eve can obtain from single photons is upper-bounded by \cite{holevo1973bounds},
\begin{eqnarray}
I_{E,1} \le \mathop {\max }\limits_{\{ U\} } \{ S({\rho ^{AE}}) - \frac{{S(\rho _0^E) + S(\rho _1^E)}}{2}\},
\end{eqnarray}
where $ S(\rho )$ is the von Neumann entropy.
Furthermore, $\rho _0^E$ and $\rho _1^E$ only differ from $\rho  \otimes \left| \varepsilon  \right\rangle \left\langle \varepsilon  \right|$ by some unitary transformations,
therefore,
\begin{eqnarray}
S(\rho _0^E) = S(\rho _1^E) = 1.
\end{eqnarray}
$S(\rho^{AE})$ can be obtained by the Gram matrix representation\cite{jozsa2000distinguishability}.
Without loss of generality, the effect of the unitary operation may be represented as,
\begin{eqnarray}
\begin{array}{c}
U\left| 0 \right\rangle \left| \varepsilon  \right\rangle  = \left| 0 \right\rangle \left| {{\varepsilon _{00}}} \right\rangle  + \left| 1 \right\rangle \left| {{\varepsilon _{01}}} \right\rangle  = \left| {{\varphi _1}} \right\rangle, \\
U\left| 1 \right\rangle \left| \varepsilon  \right\rangle  = \left| 0 \right\rangle \left| {{\varepsilon _{10}}} \right\rangle  + \left| 1 \right\rangle \left| {{\varepsilon _{11}}} \right\rangle  = \left| {{\varphi _2}} \right\rangle, \\
\sigma_ZU\left| 0 \right\rangle \left| \varepsilon  \right\rangle  = \left| 0 \right\rangle \left| {{\varepsilon _{00}}} \right\rangle  - \left| 1 \right\rangle \left| {{\varepsilon _{01}}} \right\rangle  = \left| {{\varphi _3}} \right\rangle, \\
\sigma_ZU\left| 1 \right\rangle \left| \varepsilon  \right\rangle  = \left| 0 \right\rangle \left| {{\varepsilon _{10}}} \right\rangle  - \left| 1 \right\rangle \left| {{\varepsilon _{11}}} \right\rangle  = \left| {{\varphi _4}} \right\rangle.
\end{array}
\end{eqnarray}
The corresponding Gram matrix of ${\rho ^{AE}}$ can be written as
\begin{eqnarray}
{G_{ij}} = \frac{1}{4}\left\langle {{\varphi _i}|{\varphi _j}} \right\rangle .
\end{eqnarray}
The eigenvalues of $G$ can be easily obtained as,
\begin{eqnarray}
\lambda_{1,2,3,4}  = \frac{{1 \pm {\gamma _1} \pm {\gamma _2}}}{4},
\end{eqnarray}
where ${\gamma _1} = \alpha  + \beta $, ${\gamma _2} = \sqrt {{{(\alpha  - \beta )}^2} + 4\delta {\delta ^*}}$,  $\alpha  = 1/2 - \left\langle {{{\varepsilon _{01}}}} \mathrel{\left | {\vphantom {{{\varepsilon _{01}}} {{\varepsilon _{01}}}}}\right.\kern-\nulldelimiterspace}{{{\varepsilon _{01}}}} \right\rangle $, $\beta  = \left\langle {{{\varepsilon _{10}}}}
\mathrel{\left | {\vphantom {{{\varepsilon _{10}}} {{\varepsilon _{10}}}}}\right. \kern-\nulldelimiterspace}{{{\varepsilon _{10}}}} \right\rangle  - 1/2$, and $\delta  = \left\langle {{{\varepsilon _{00}}}}\mathrel{\left | {\vphantom {{{\varepsilon _{00}}} {{\varepsilon _{10}}}}}\right.\kern-\nulldelimiterspace}{{{\varepsilon _{10}}}} \right\rangle $.

Obviously, $ S({\rho ^{AE}})$ is monotonically decreasing with $\gamma _1$ and $\gamma _2$.
Therefore, it has a maximum when ${\gamma _1} = \alpha  + \beta  = \left\langle {{{\varepsilon _{10}}}}
\mathrel{\left | {\vphantom {{{\varepsilon _{10}}} {{\varepsilon _{10}}}}}
	\right. \kern-\nulldelimiterspace}
{{{\varepsilon _{10}}}} \right\rangle  - \left\langle {{{\varepsilon _{01}}}}
\mathrel{\left | {\vphantom {{{\varepsilon _{01}}} {{\varepsilon _{01}}}}}
	\right. \kern-\nulldelimiterspace}
{{{\varepsilon _{01}}}} \right\rangle  = 0$, ${\gamma _2} = \sqrt {{{(\alpha  - \beta )}^2} + 4\delta {\delta ^*}}  \ge \left| {\alpha  - \beta } \right| = 1 - 2 \cdot \left\langle {{{\varepsilon _{01}}}}
\mathrel{\left | {\vphantom {{{\varepsilon _{01}}} {{\varepsilon _{01}}}}}
	\right. \kern-\nulldelimiterspace}
{{{\varepsilon _{01}}}} \right\rangle $. 

The BER of single photons measured in the Z-basis is ${e_{m,1}} = \left\langle {{{\varepsilon _{01}}}}
\mathrel{\left | {\vphantom {{{\varepsilon _{01}}} {{\varepsilon _{01}}}}}
	\right. \kern-\nulldelimiterspace}
{{{\varepsilon _{01}}}} \right\rangle $.
Then, the upper bound of $I_{E,1}$ can be written as 
\begin{eqnarray}
I_{E,1} = h({e_{m,1}}) 
\end{eqnarray}
where the value of ${e_{m,1}}$ can be upper-bounded by the decoy state methods for the states in Z-basis at Alice's end.
    
Thus, the SKR is
\begin{eqnarray}
    R_s = Q \cdot [\Delta_1-\Delta_1\cdot h(e_{m,1})+\Delta_0-f\cdot h(e)].
\end{eqnarray}
Clearly $R_s$ will decrease soonly or even reach zero when $e$ and ${e_{m,1}}$ increase. Therefore, a high ratio of the maximum tolerable error rate to the experimental error rate reflects a high robustness to environmental noise.

\subsection{Performance analysis}
\begin{figure}[h!]
\centering
	\includegraphics[width=7cm]{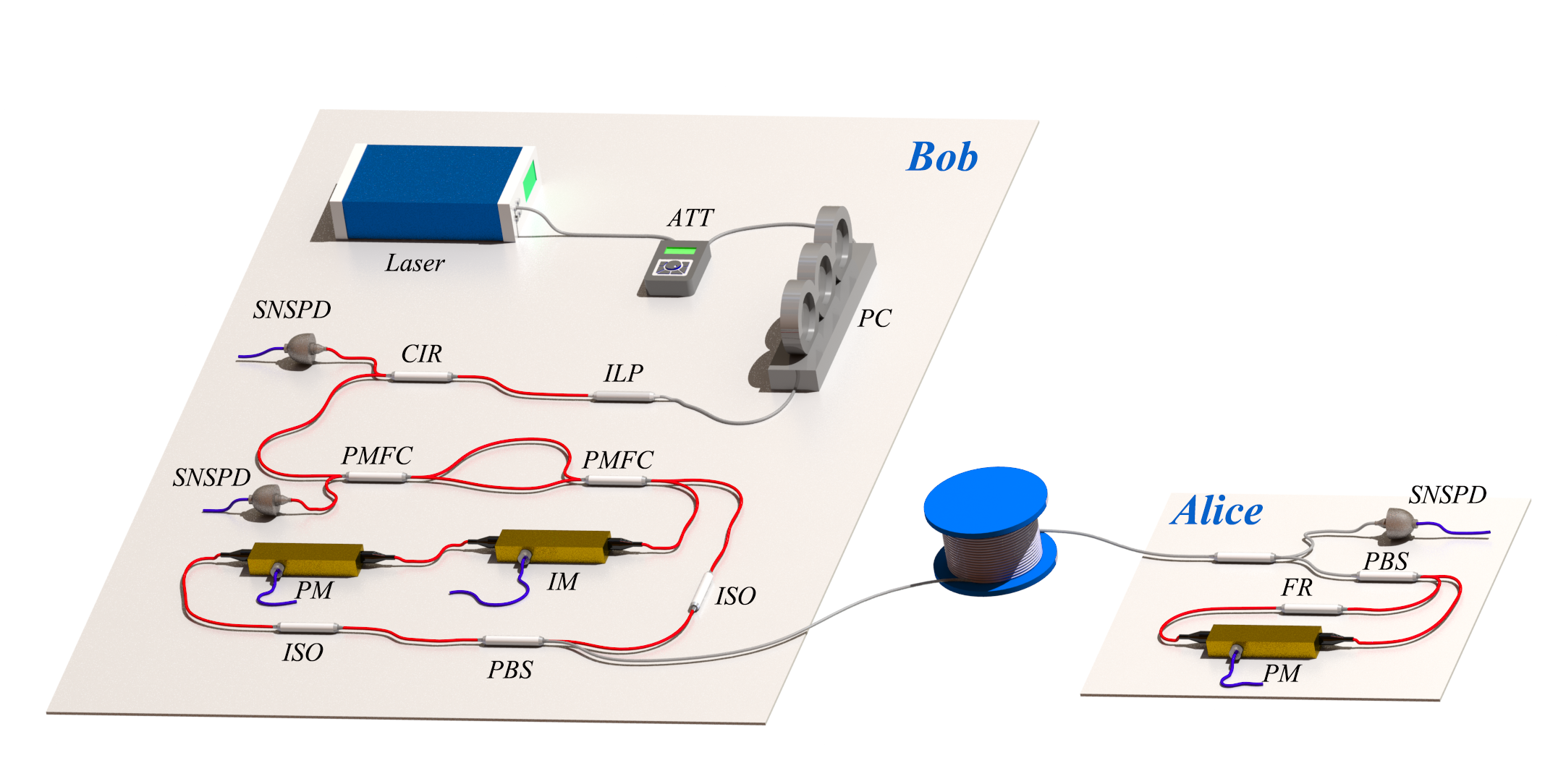}
	\caption{Experiment setup. Laser: 1550 nm with a systematic pulse-repetition frequency of 50 MHz; ATT: attenuator, used for attenuating the laser and simulating the decoy states instead of another intensity modulator; PC: polarisation controller; ILP: in-line polarizer; CIR: optical circulator; PBS: polarisation beam splitter; PMFC: polarisation maintaining filter coupler; PM: 2.2dB insertion loss commercial lithium niobate phase modulator; IM: intensity modulator with an extinction ratio greater than 40 dB; ISO: isolator; FR: Faraday rotator; SNSPD: superconducting nanowire single-photon detector with 85 \% detection efficiency, lower than 100 Hz dark count rate and 15 ns reset time. The length of delay is 2 m.}
	\label{setup}
\end{figure}
The experimental set-up is shown in Figure \ref{setup}.
The mean photon numbers of the signal state and decoy state are $u=0.6$ and $v=0.2$, respectively, with a 50 MHz repetition rate. Two subpulses pass through the same interferometer, so that the phase difference of the two pulses is stable. To control the polarisation, a ring with a Faraday rotator was designed to replace the Faraday mirror \cite{sun2010quantum}, and all the modulators were polarisation-sensitive, which makes the system more stable. In particular, the subpulses only go through the modulators once and there was no crosstalk between the forward and backward channels. This design allows the use of a repetition rate greater than 1 GHz.

\begin{table*}
	\caption{\bf Experimental parameters and performance}
	\label{parameter}
	\begin{tabular}{c c c c c c c c c}
		\hline
		\hline
		Length&$Q$& $\Delta_1$ & $\Delta_0$ & $e_{m,1}$ & $e$ & Block size & $f$ & SKR \\
		\hline
		50.4 km&$1.21\times 10^{-4}$ & $0.4797$ & $0.0017$ & $0.0008$ &$0.0064$& $3\times 10^{7}$ &$1.2$ & $4.956\times 10^{-5}$\\
		\hline
		\hline
	\end{tabular}
\end{table*}

In an 80-minute test, the BER in the X-basis is 0.64 \%, 
At Bob's end, the gains of signal state, decoy state and vacuum state in the X-basis are $Q_u^X=1.21\times 10^{-4}$, $Q_v^X=3.96\times 10^{-5}$ and $Q_0^X=3.67\times 10^{-7}$, respectively.
We can estimate the fraction of detection events from the single-photon and vacuum state as,
\begin{eqnarray}
  	\begin{array}{c}
		{\Delta _0} = {Q_0^X}/{Q_u^X} \cdot {e^{ - u}} = 0.0017, \\
		{\Delta _1} = {u}/{Q_u^X} \cdot {e^{ - u}}\cdot {Y^{L,X}} = 0.4797,
	\end{array}
\end{eqnarray}
where ${Y^{L,X}} = u \cdot [Q_v^X{e^v} - {v^2}/{u^2} \cdot Q_u^X{e^u} - ({u^2} - {v^2})/{u^2} \cdot Q_0^X]/(uv - {v^2})$.

At Alice's end, the BER of signal state and decoy state in the Z-basis are both 0.06 \%.
The gain of signal state, decoy state and vacuum state in the Z-basis are $Q_u^Z=1.97\times 10^{-2}$, $Q_v^Z=6.32\times 10^{-3}$ and $Q_0^Z=3.81\times 10^{-7}$, respectively.
We can estimate the BER for single photons in the Z-basis as,
\begin{eqnarray}
    e_{m,1} = \frac{{{E_v}Q_v^Z{e^v}}-\frac{1}{2}Q_0^Z}{{v\cdot {Y^{L,Z}}}} = 0.08\%,
\end{eqnarray}
where ${Y^{L,Z}} = u \cdot [Q_v^Z{e^v} - {v^2}/{u^2} \cdot Q_u^Z{e^u} - ({u^2} - {v^2})/{u^2} \cdot Q_0^Z]/(uv - {v^2})$.
 The block size is $3\times 10^{7}$, and we take the reconciliation efficiency $f = 1.2 $, so that the SKR is $4.956\times 10^{-5}$, which is about 2.5 kbps. The actual SKR will decrease if the modification of the finite-key \cite{tomamichel2012tight,2013Experimental,2015Finite} and the occupation of monitoring and decoy states are considered.
Table \ref{parameter} shows the experimental parameters and performance indices.

\begin{figure}[htbp]
\centering
	\includegraphics[width=7cm]{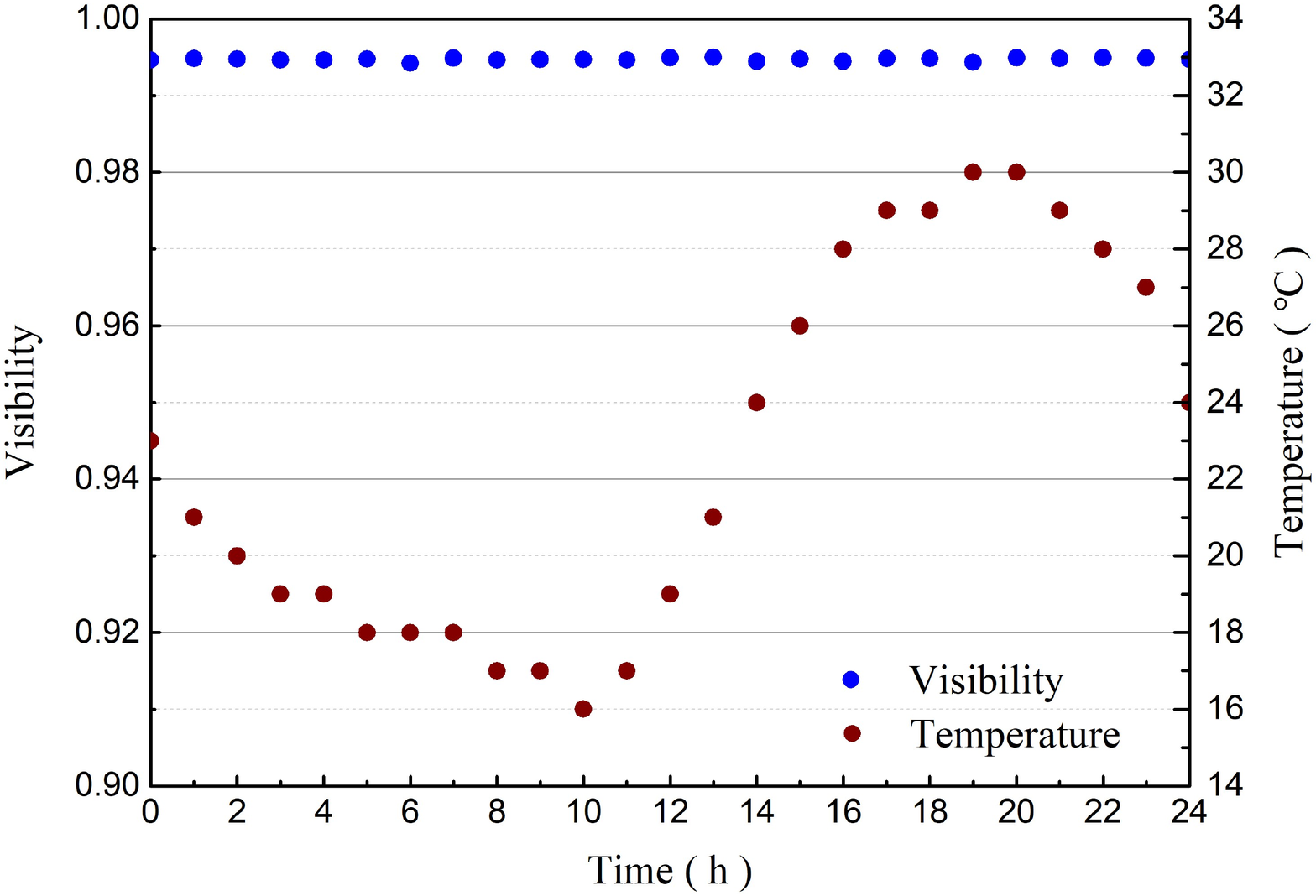}
	\caption{System stability over 24-hour environment change for a length of 50.4 km with visibility of $0.9947\pm 0.0002$.}
	\label{visibility}
\end{figure}

In order to eliminate the influence of the dark count rate (DCR) and modulation errors, we use stronger pulses and only ecode 0 to test the performance. Figure \ref{visibility} shows the visibility being maintained at over 0.994 for more than 24 hours. To imitate significant environmental changes, the entire system is placed next to an open window with a randomly changing polarisation controller connected to the quantum channel.
In practice, more than 99\% of error bits are caused by Rayleigh backscattering distributed over the entire time domain, which means there is nearly no mismatching or phase drift of the interferometer in our system. For the signal state in the X-basis, the DCR contributes to about 0.04\% of the error rate. The Rayleigh elastic backscattering actually contributes to the error rate twice as much compared to the 24-hour test, which is equal to 0.53\%. The  remaining 0.07\% is due to electronic control errors. For the state in the Z-basis, the intensity modulator with an extinction ratio greater than 40dB contributes 0.01\% to the error rate. Thus, the total error rate will easily be lower than 0.02\% using an accurate temperature controller.

\begin{figure}[htbp]
\centering
	\includegraphics[width=7cm]{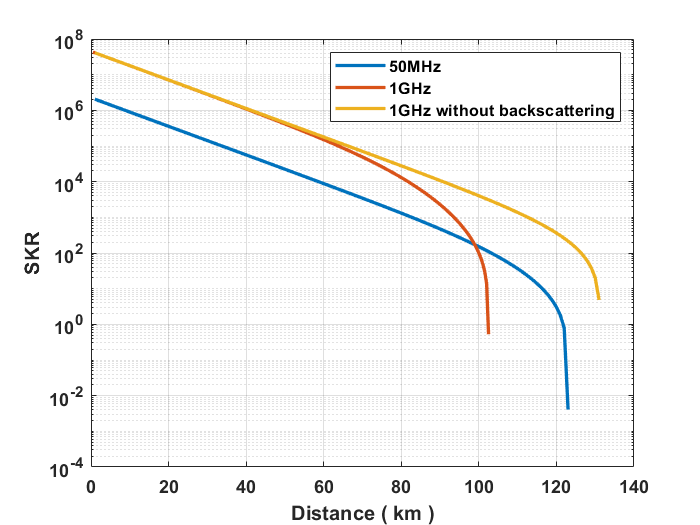}
	\caption{The SKR for 50 MHz, 1 GHz repetition rates and 1 GHz repetition rate without backscattering. We assume $\beta=-40.5dB$, $DCR=5\times 10^{-8}$, the loss of fibre is $0.2dB/km$, the total loss of encoding and decoding section is $4dB$ and $5dB$, respectively.}
	\label{simulation1}
\end{figure}

Here, we simulate the SKR of our scheme at repetition rates of 50 MHz and 1 GHz, as well as 1 GHz without backscattering, as shown in Figure \ref{simulation1}. For simplification and optimisation, we choose the mean photon number $u$ of the signal states to maximise the SKR, and the decoy state can perfectly estimate all the parameters. A 90:10 coupler at Alice’s end and much weaker states in the Z-basis were used to increase the SKR. 
When the loss or the repetion rate increases ,the Rayleigh backscattering causes significant difficulties for two-way QKD because the ratio of the backscattered power to the entering power is nearly fixed at the value $\beta$. There are different values $\beta$ for different fibres, with ordinary fibres typically having $\beta$ values of approximately $-40.5dB$ \cite{2004Backscattering}. Some special fibre manufacturing processes can reduce Rayleigh backscattering, such as increasing the core diameter and changing the doping of materials. However, when the repetition rate increases, the influence of backscattering is still greater than the DCR. When Rayleigh backscattering contributes excessively to the BER, another fibre must be used as the backward channel to eliminate the influence. Under this circumstance, a polarisation controller is needed before decoding to maintain stability and high gain at Bob’s end.

\section{Discussion}
Although this protocol has been proved to be theoretically secure, there are still possible loopholes need to be considered in practical applications. In order to estimate the value of ${e_{m,1}}$ accurately, the detection efficiency of Alice's detector for all the photons entering Alice's end should be equal. Therefore, depending on the different measurement devices, different modifications are required for eliminating the possible loopholes at Alice's end. For example, if the detection efficiency is sensitive to the polarisation of photons, a randomly changing polarisation controller or a polarisation beam splitter with one more detector will be possible solutions; When the detector's reset time covers some time windows, the corresponding parts of raw key can be discarded. Trojan horse attacks at Bob's end should also be strictly monitoring. By adding an attenuator after the modulators, the information leakage can be limited \cite{2015Practical}. Besides the device-independent QKD protocols, the above two experimental vulnerabilities are also found in typical QKD protocols. This new design eliminates theoretical vulnerabilities while retaining the advantages of high robustness, simplicity and flexibility of typical plug-and-play scheme.

\section{Conclusion}
In conclusion, the long-term stable operation and the extremely low BER without active feedback show the high robustness of the system. The loophole-free plug-and-play design with an arm-length-insensitive interferometer provides great security and flexibility for a range of applications. Indeed, the achievable distance of the two-way QKD system can be roughly half of the typical one-way QKD schemes when the same SKR is required. However, different schemes will be used in different situations in the future construction of quantum communication networks. Practical applications are always the trade-off between reliability, flexibility and capacity. Our scheme realises a combination of high reliability and flexibility at the expense of the maximum feasible distance.

\section*{Funding.}
This work was supported by the National Key Research and Development Program of China under Grant No.2017YFA0303700 and the National Natural Science Foundation of China under Grant No.11974205, the Key Research and Development Program of Guangdong province (2018B030325002) and Beijing Advanced Innovation Center for Future Chip (ICFC).

\section*{Acknowledgment.}
The authors thank Prof. Qiang Zhou for helpful discussions.

\section*{Data availability}
Data underlying the results presented in this paper are not publicly available at this time but may be obtained from the authors upon reasonable request.

\section*{Disclosures.}

The authors declare no conflicts of interest.


\bibliography{Ref}






\end{document}